\documentclass[sigconf,10pt]{acmart}
\AtBeginDocument{%
  }

\setcopyright{none}
\renewcommand\footnotetextcopyrightpermission[1]{}  

\usepackage{enumitem}

\usepackage{pbalance}

\usepackage{listings}
\usepackage{xcolor}

\begin{document}

\title{Distributed AI Platform for the 6G RAN}

\author[G. Ananthanarayanan, M. Balkwill, X. Foukas, Z. Lai, B. Radunovic, C. Settle, Y. Zhang]{Ganesh Ananthanarayanan, Matthew Balkwill, Xenofon Foukas, Zhihua Lai, Bozidar Radunovic, Connor Settle, Yongguang Zhang}
\email{{ga, Matthew.Balkwill, xefouk, zhihualai, bozidar, connorsettle, ygz}@microsoft.com}
\affiliation{%
  \institution{Microsoft Research}
  \city{}
  \state{}
  \country{}
}

\begin{abstract}
Cellular Radio Access Networks (RANs) are rapidly evolving towards 6G, driven by the need to reduce costs and introduce new revenue streams for operators and enterprises. 
In this context, AI emerges as a key enabler in solving complex RAN problems spanning both the management and application domains.
Unfortunately, and despite the undeniable promise of AI, several practical challenges still remain, hindering the widespread adoption of AI applications in the RAN space.
In this work, we attempt to shed light to these challenges and argue that existing approaches in addressing them are inadequate for realizing the vision of a truly AI-native 6G network.
We propose a distributed AI platform architecture, tailored to the needs of an AI-native RAN.
\end{abstract}

\maketitle

\pagestyle{plain}

\section{Introduction}

\begin{figure}[t]
	\centering
	\includegraphics[width=0.90\columnwidth]{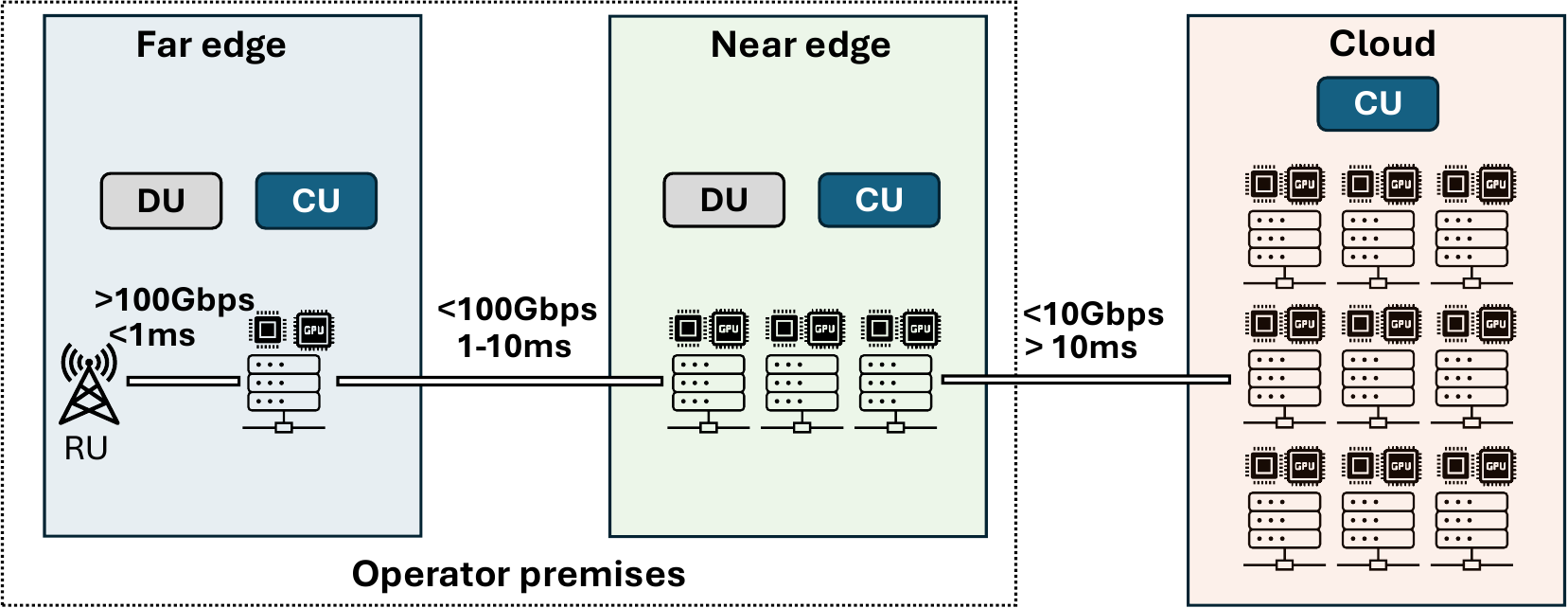}
	\caption{\bf High-level 5G RAN overview. The RAN infrastructure capabilities can vary depending on the location.}
	\label{fig:tradeoffs}
    \vspace{-2mm}
\end{figure}

In 5G, RAN is undergoing a paradigm shift, disaggregating monolithic base stations into separate components (CU, DU, and RU) and containerizing them on commodity hardware/software platforms across edges and cloud (Fig.~\ref{fig:tradeoffs}).
This enables open interfaces for RIC-based applications, accelerating innovation and facilitating new capacity-boosting technologies like Massive MIMO, mmWaves, and densification.
6G will further address long-standing problems (e.g., radio resource management, mobility, energy savings) and introduce transformative uses (e.g., joint communication-sensing, security, slicing).\looseness=-1

The AI revolution, as catalyzed by the current generative AI phenomenon, has motivated the telecom industry to see AI as an ideal fit for RAN problems like signal classification, traffic prediction, and optimal scheduling.
The next-gen mobile networks, including 6G, will be ``AI-native''.
AI-RAN Alliance~\cite{airan-alliance}, a telecom industry's AI initiative, has defined three promising domains of AI use cases in the RAN:

{\bf (1) AI-for-RAN} -- use AI to optimize RAN performance.
    For example, higher frequencies, cell density, and larger antenna arrays will increase the search space in scheduling with AI-based solutions~\cite{bartsiokas2022ml}. Infrastructure management like predictive maintenance of RAN sites, vRAN troubleshooting~\cite{sun2024spotlight}, and energy savings~\cite{kundu2024towards, singh2021energy, kalia2025towards} will all benefit from AI.

{\bf (2) AI-and-RAN} -- share compute (CPU/GPU) between RAN and AI apps.
The vRAN is largely underutilized ($<50\%$)~\cite{nvidia_vran} due to overprovisioning and traffic inbalance~\cite{foukas2021concordia}, but AI can be leveraged to achieve sharing without violating the RAN's realtimeness~\cite{foukas2021concordia,schiavo2024yinyangran, foukas2025future}.

{\bf (3) AI-on-RAN} -- use RAN infrastructure for AI apps.
    For example, localization from channel data for tracking/autonomy~\cite{trevlakis2023localization}, context-aware security~\cite{chorti2022context}, and latency-critical video analytics~\cite{zhang2024vulcan} will benefit from interfacing with RAN.

Despite these promises, challenges persist.
First, we need to collect data from vastly and geo-distributed RAN sites for accurate AI models.
Second, AI-RAN use cases vary in compute, latency, and privacy needs, complicating edge-cloud orchestration. The goal of this research is to design a distributed AI platform for 6G RAN.
We highlight the issues, arguing static approaches fall short for AI-native networks.
We propose a distributed AI architecture to address these problems.

\section{Background}

\subsection{Canonical application examples}
\label{subsec:examples}

\begin{figure}[t]
	\centering
	\includegraphics[width=0.90\columnwidth]{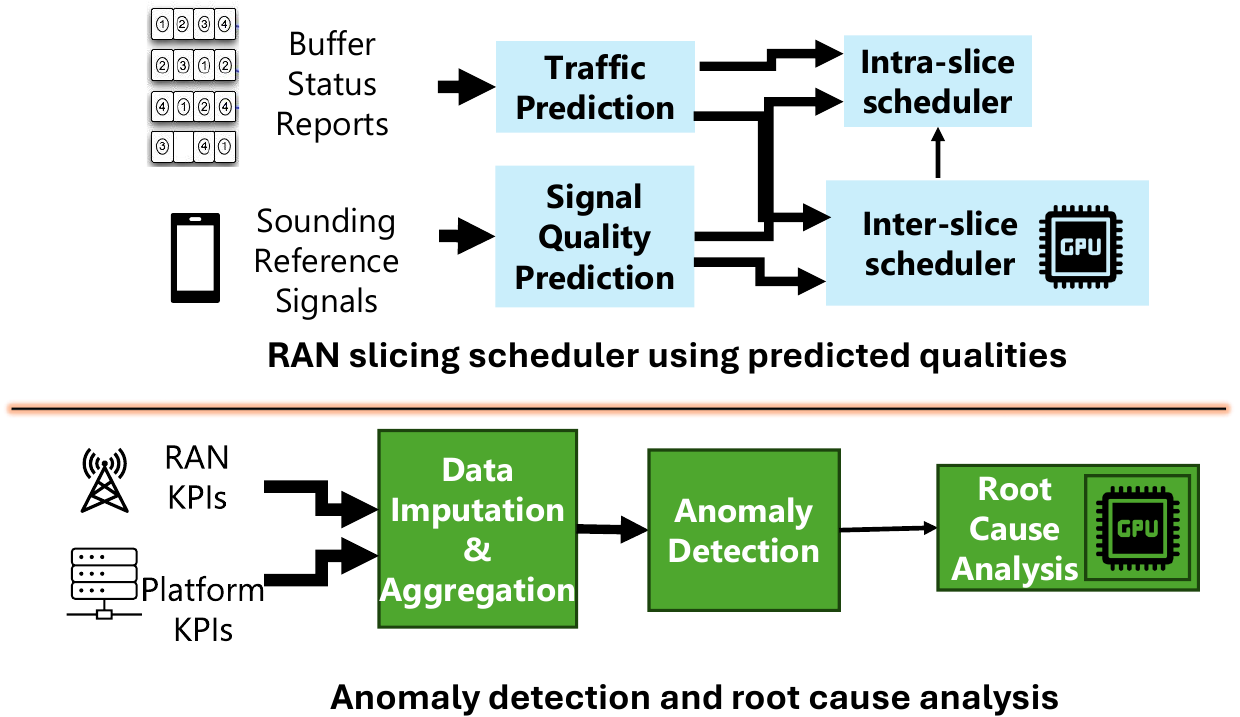}
	\caption{\bf Graphs of AI/ML models for RAN applications}
	\label{fig:ai_models}
    \vspace{-2mm}
\end{figure}

To help with highlighting the need for a distributed AI RAN platform, we define two examples (Fig.~\ref{fig:ai_models}), which we use as a reference for the remainder of this work.

\noindent\textbf{RAN slicing scheduler --}
A scheduler allocates radio resources across network slices (inter-slice scheduling) and across users of the same slice (intra-slice scheduling). 
Both schedulers use the traffic demand information provided by the phones through buffer status reports, to try and predict the traffic load for the upcoming period~\cite{slicing,foukas2017orion}. 
They also use physical layer sounding reference signals to predict the users’ signal quality.
The inter-slice scheduler makes scheduling decisions at coarse granularities (e.g., seconds) and feeds its decisions to the real-time intra-slice scheduler. 
This example illustrates both AI-on-RAN (predicting load using buffer status reports) as well as AI-for-RAN (scheduling decisions).

\noindent\textbf{Anomaly detection and root cause analysis --} 
A service management and orchestration framework tries to detect RAN-related anomalies and identify their root cause for potential mitigations towards AI-for-RAN.
Since the anomalies can originate both from the RAN and the platform, the application collects data from both sources~\cite{sun2024spotlight}.
It also applies imputation and aggregation techniques, to deal with noisy/missing data and to reduce their volume.
It then processes the collected statistics, to detect if an anomaly is present, and if so, it attempts to localize it and detect its root cause.

From these examples, we see that a typical AI RAN application, rather than being a monolith, may consist of a sequence of blocks.
Each block can do its own data pre-processing and can include an ML model.
Those blocks are chained together into a graph, where the output of one model becomes the input to the next.
Typically the traffic volume between components reduces from left to right, as more data gets processed and aggregated.\looseness=-1

\subsection{Distributed edge infrastructure}

The RAN infrastructure is typically deployed across a distribution of far and near edges, extending to the cloud (Fig.\ref{fig:tradeoffs}). As we move from the far edge to the cloud, more compute resources become available for the AI RAN applications. 
However, choosing the cloud for deployment (e.g., due to the presence of GPUs) is not always the right choice.
The abundance of resources comes at the expense of reduced bandwidth and higher latency. 
This can be crucial factors for applications such as the example anomaly detector, which requires large volumes of input data, or the RAN slicing scheduler, which is latency-sensitive in its allocation decisions. 
\section{Challenges to build AI RAN applications}

We now discuss the challenges of deploying AI RAN applications, that make the case for a distributed AI RAN platform.\looseness=-1 

\subsection{Data collection}

Different applications require different feature sets, that might have heterogeneous characteristics in terms of type, time granularity, etc.
For example, the radio resource scheduling application might require real-time data from the RAN network functions, while the anomaly detection might need to combine aggregate data from both the RAN and the platform in a time windows of seconds.
Similarly, the anomaly detection application might rely on the inter-arrival time between IQ samples, while the scheduler might require the actual raw IQ samples carrying the sounding reference signals.

The heterogeneity in the input features of AI applications is what makes the data collection process challenging.
Exposing raw data from all possible data sources is not a viable option, as it would lead to a huge volume of data that one would have to process, store and transmit. 
For example, capturing all the raw IQ samples from the physical layer of the RAN translates into several gigabits of traffic per second, even for a single base station with four antennas.
Similarly, capturing all the CPU scheduling events of a server, in order to detect if a platform anomaly due to CPU interference is present, would result in the collection of millions of events per second.

The current standard practice to bypass this roadblock is to expose a set of coarse-grained data sources, that are applicable to a wide range of use cases. 
These data sources are exposed through static APIs specified by standardization bodies like 3GPP and O-RAN.
For instance, O-RAN defines the E2 interface for the collection of pre-defined RAN KPIs, and the O2 interface for the collection of platform data.
 Any change to a data source requires the standardization body consensus, which can be a very long process.
As such, AI RAN applications end up being developed as an afterthought, subject to the available data sources, rather than in a truly AI-native way, where the data collection is application-driven.


\subsection{RAN AI application orchestration}

The disaggregated nature of the RAN means that its network functions might be deployed across the edges and the cloud.
This raises a fundamental question as to what should be the location in which the blocks of AI applications should reside.
Answering this question is far from straightforward.
It involves matching the applications' requirements to the capabilities of the infrastructure, which can vary in terms of compute resources, network bandwidth, latency (Fig.~\ref{fig:tradeoffs}) and privacy.

The many constraints and trade-offs make the deployment of AI RAN applications a major challenge. 
First, a developer needs to understand the underlying constraints which may differ from one deployment to another.
Second, they need to carefully choose what data to collect and where to place the application blocks, to maintain high accuracy, while respecting the constraints.
Third, deployed applications will naturally have competing requests for the same resources, meaning that it falls on the developer to design their application with placement flexibility in mind. 
Finally, AI RAN applications have to coexist with other AI applications that also execute on the edge, such as video analytics and self-driving cars~\cite{zhang2024vulcan}.
In the absence of a structured framework, AI RAN applications have to be developed with ad hoc programming interfaces and manually distributed by developers. 
\section{Distributed AI platform for RAN}
\label{sec:platform}

\begin{figure}[t]
	\centering
	\includegraphics[width=0.9\columnwidth]{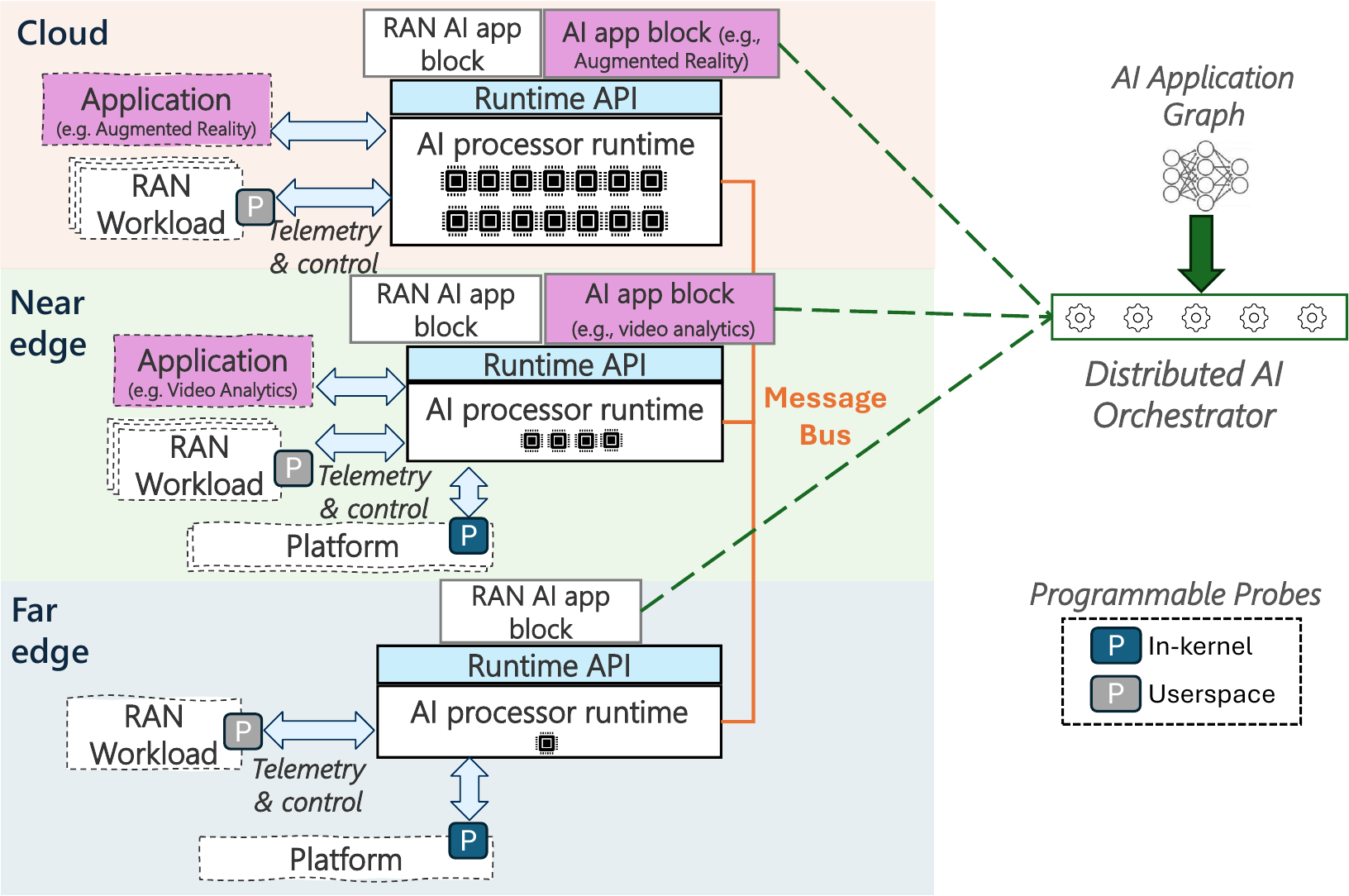}
	\caption{\bf Distributed AI platform architecture.}
	\label{fig:ai_platform}
    \vspace{-2mm}
\end{figure}

We present our vision for distributed AI-native RAN platform, with the architecture illustrated in 
Fig.~\ref{fig:ai_platform}. It builds on top of three components; i) programmable probes for the flexible collection of data, ii) AI processor runtimes for the deployment of AI applications across the distributed compute fabric, and iii) an orchestrator for the coordination of the platform. 

\noindent\textbf{Programmable probes for dynamic data collection --}
To allow developers to define optimal feature sets for their AI applications, we propose the use of dynamic probes for the platform (i.e., OS kernel) and the userspace (i.e., RAN network functions) to programmatically collect data. 
Through the probes, a developer could write small pieces of code to access raw events and data structures and summarize them in a custom way.
This would expose the right features for training and inference, while minimizing the volume of generated data.
For example, a developer could leverage a probe to access the raw IQ samples of the base station to feed them directly to the application, like in the example of the RAN slicing scheduler.
Alternatively, they could pre-process them and export a derived KPI, like in the case of the anomaly detection application. 
The same approach could also be used to capture platform data (e.g., capture the incoming TCP packets and calculate the average inter-packet delay).

While this approach provides flexibility in tailoring the data collection process to the AI application's requirements, it also introduces safety concerns (e.g., illegal memory accesses).
For this, we propose to use probes based on the eBPF technology, which has recently caught the attention of the telco industry~\cite{foukas2023taking, soldani2023ebpf}.
eBPF allows the injection of code to instrumentation points, subject to a static verification process that guarantees the safety of the injected code.
While eBPF originates in the Linux kernel, recent advances~\cite{foukas2023taking} have expanded its use to userspace RAN applications. 
\newline

\noindent\textbf{AI processor runtime --}
Considering the constraints and trade-offs that characterize the various locations of the infrastructure, an AI-native platform should allow the seamless deployment of AI applications to the most suitable location, without the developer having to worry about providing location-specific flavors.
In the proposed architecture, this is achieved through the \emph{AI processor runtime}.
The runtime can be deployed at any location where AI applications are expected to run and introduces an API for its interactions with the applications.
The runtime should provide the following functionalities:

\emph{-- Data ingestion and control:} It should enable the ingestion of data captured through the local programmable probes and their exposure to applications. It should also enable the issuing of  API calls towards the RAN functions and the platform for closed loop control.

\emph{-- Data exchange:} It should allow the exchange of data streams among the AI application blocks deployed in different locations through a common message bus. 
Similarly, it should enable the issuing of RPCs for applying control decisions.

\emph{-- Execution environment:} It should implement the process of running inference or training tasks, by abstracting the underlying compute resources (e.g., CPUs and GPUs) and exposing them to both RAN and non-RAN AI applications.
\newline
\emph{-- Life-cycle management:} It should provide a standard interface to deploy, update, and remove AI applications, as well as to monitor their performance and resource utilization. 

As long as it provides the functionalities described above, we eschew making an implementation prescription because we believe that the framework should be extensible to include existing and future runtime environments and messaging technologies, thus not restraining AI developers. 
As an example, one could consider using Docker containers combined with a WebAssembly (WASM) runtime as a highly portable and sandboxed execution environment, while the message bus implementation could be based on REST or gRPC calls. 

\begin{figure}[t]
	\centering
	\includegraphics[width=0.9\columnwidth]{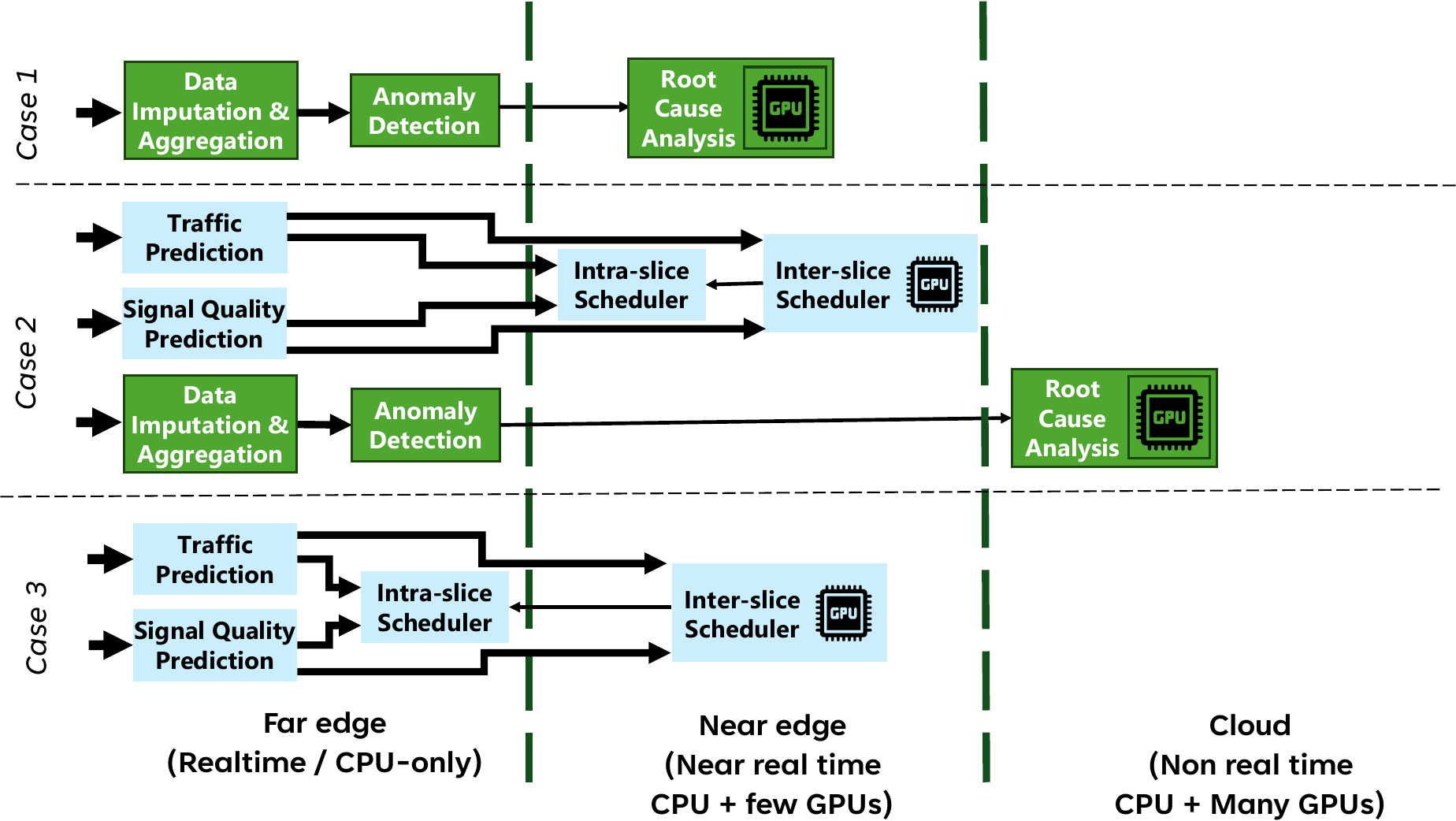}
	\caption{\bf Distributed deployment scenarios for orchestrated AI applications.}
	\label{fig:model_split}
    \vspace{-2mm}
\end{figure}

\noindent\textbf{Orchestrator --}
The distributed AI framework is overseen by an orchestrator (Fig~\ref{fig:ai_platform}) that is responsible for the placement of the AI applications across the AI processor runtimes.
The orchestrator allows for dynamic addition or removal of applications, and also handles dynamic changes to the available resources (compute and network). 
It also allows to plug in diverse policies that trade off between compute at the different edges and the network load between the edges.

Applications are exposed to the orchestrator in the form of blocks in a graph.
The blocks are characterized compute, memory, network and latency requirements, as well as other constraints (e.g., the locations where the block is not allowed to run due to privacy concerns).
The orchestrator takes into account both the developer requirements and the capabilities of the infrastructure and places the blocks to the AI processor runtimes that maximize the overall utility of the platform.
Unique to AI workloads are {\em inference parameters} that influence resource demand \cite{zhang2024vulcan}, and hence is a resource allocation knob for the orchestrator. 
Examples of inference parameters would be data sampling rates, or AI models with varying accuracy for anomaly detection. 
We propose the AI RAN applications to expose their parameters to the orchestrator, and design the orchestrator to dynamically change them. 

We use the example applications of Fig.~\ref{fig:ai_models} to explain the operation of the orchestrator.
We consider a far-edge location without a GPU and a limited amount of CPUs, a near-edge location with a single GPU, and a cloud location with many GPUs.
Under this setup, we consider the three cases of Fig.~\ref{fig:model_split}.
In the first case, we install the anomaly detection application. 
Given that the root cause analysis model requires a GPU, the orchestrator places it at the near edge and the rest of the applications' blocks at the far edge, minimizing the amount of outgoing traffic.
In the second case, we install the RAN slicing scheduler application in addition to the anomaly detector.
The latency-sensitive inter-slice scheduler requires a GPU, however the one available at the near edge is already reserved.
Therefore, the orchestrator migrates the root cause analysis model to the cloud, since it is not latency sensitive, and deploys the inter-slice scheduler to the near edge.
The latency-tolerant intra-slice scheduler is also deployed at the near edge, since there is not enough CPU capacity at the (preferable) far edge location.
Finally, in the third case, we illustrate how the orchestrator handles the completion of the anomaly detector and reschedules the radio resource scheduler. 
With the anomaly detector removed, CPU resources are freed up at the far edge, so the orchestrator migrates the intra-slice scheduler there, reducing the traffic sent to the near edge.

\section{An efficient far edge AI processor runtime}
\label{sec:decima}

We advocate that a highly optimized far edge runtime is required to host real-time AI applications (e.g. the RAN slicing scheduler), as has been demonstrated in recent research works (e.g.,~\cite{ko2024edgeric}), and as envisaged in the O-RAN concept of dApps~\cite{d2022dapps}.
As such, it needs to have sub-millisecond reaction times.
Furthermore, the far edge is resource constrained, with only a small fraction of CPUs  available for AI applications, and with a small or no GPU present. 
Our proposed design is illustrated in Fig.~\ref{fig:decima_processor} and has the following characteristics:

\begin{figure}[t]
	\centering
	\includegraphics[width=0.9\columnwidth]{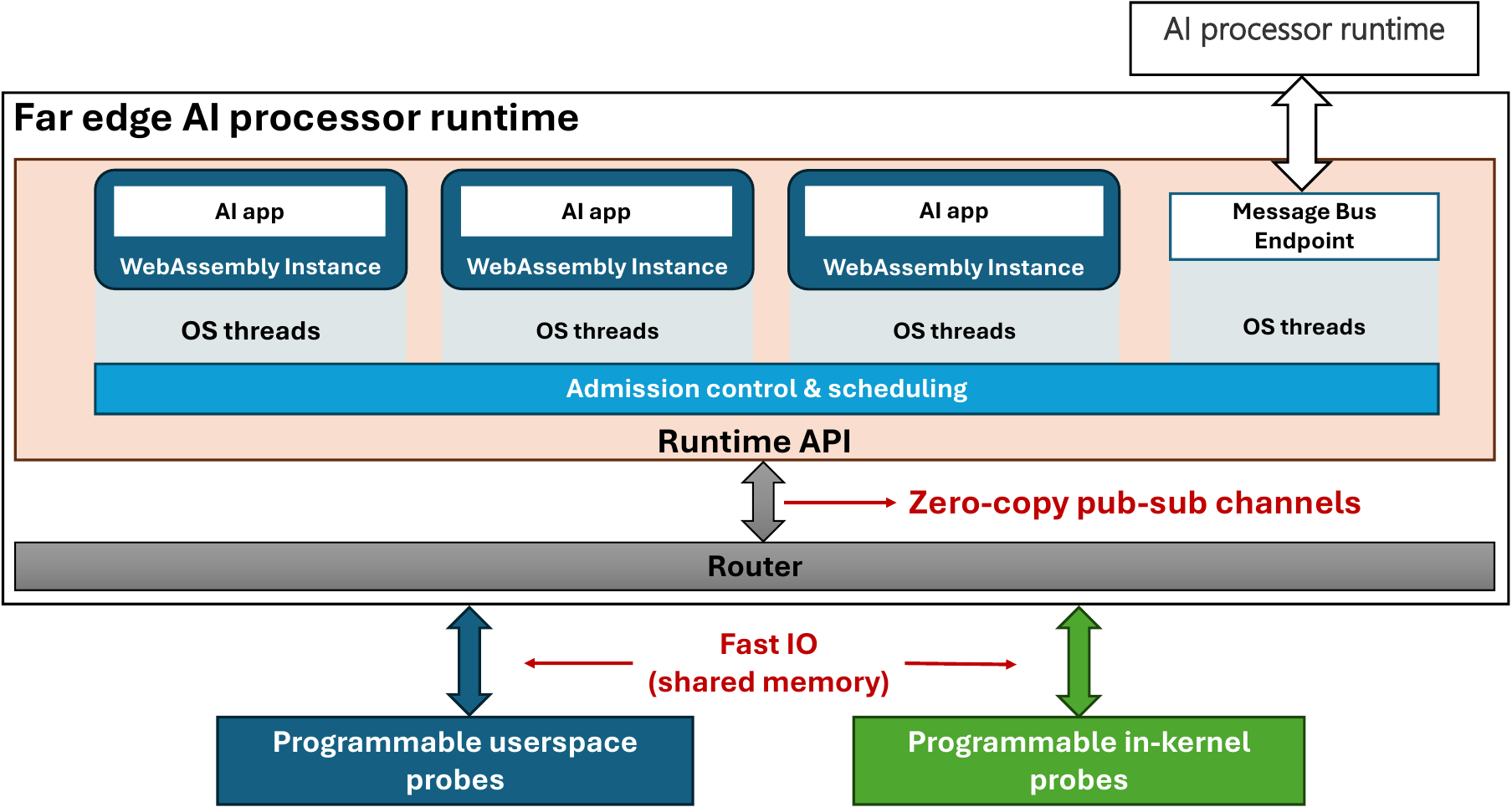}
	\caption{\bf Architecture of far edge AI processor runtime.}
	\label{fig:decima_processor}
    \vspace{-2mm}
\end{figure}

\begin{description}[style=unboxed,leftmargin=0cm]
	\item[Tight integration with the probes --] The AI processor runtime communicates with the probes through fast IO channels using shared memory and a zero-copy mechanism.
	This allows the passing of data between the probes and the AI applications with very low compute and latency overhead.
	\item[Efficient, flexible and secure execution environment -- ] We have experimented with the use of WASM as an execution environment and observed its benefits.  
	It enables the sandboxing of applications with a very small overhead, while maintaining near-native runtime performance.
	The interaction with the rest of the system for the acceleration of inference or the data collection can take place through API calls exposed by the WASM Interface (WASI). 
	For example, wasi-nn could be used for the exposure of the CPUs and the GPUs (when available). 

	\item[Admission control \& real-time resource allocation -- ] All AI applications deployed on the processor runtime need to request a fraction of the CPU time or the GPU area.
	An admission control process ensures that this request can be met, while also serving other applications.
	Once deployed, an appropriate schedule is imposed by the runtime (for example, for the CPU we leverage the Linux deadline scheduler).
\end{description}

\section{Implementation \& Evaluation}
\label{sec:eval}

To demonstrate the feasibility and benefits of our proposed architecture, we implemented a reference far edge AI processor runtime, based on the design outlined in Section~\ref{sec:decima}, and also prototyped the distributed AI orchestrator.

\subsection{Far edge AI processor runtime}
\label{sec:jrtc}

Our reference AI processor runtime is written in $\sim$5K lines of C/C++ code and is publicly available at~\cite{jrtcontroller}. 
The runtime provides both a C++ and a Python API for the development of applications, to cater the needs of developers requiring either high performance (C++) or quick integration with popular AI frameworks and libraries (Python). 
It should be noted that the current reference implementation does not support WASM for deploying applications. The applications can instead be loaded in the form of shared libraries (.so), dynamically linked at runtime through a REST-based API.
The extension to WASM applications is left as future work.

For the programmable in-kernel probes, we leveraged the popular  eBPF technology that is part of the Linux kernel~\cite{ebpf}, while for the userspace probes, we leveraged the jbpf userspace eBPF framework~\cite{jbpf, foukas2023taking}, which has been designed with lightweight RAN instrumentation in mind.
The integration of the probes with the AI processor runtime is done using zero-copy shared memory channels that are part of the jbpf framework.

\begin{figure}[t]
	\centering
	\includegraphics[width=0.7\columnwidth]{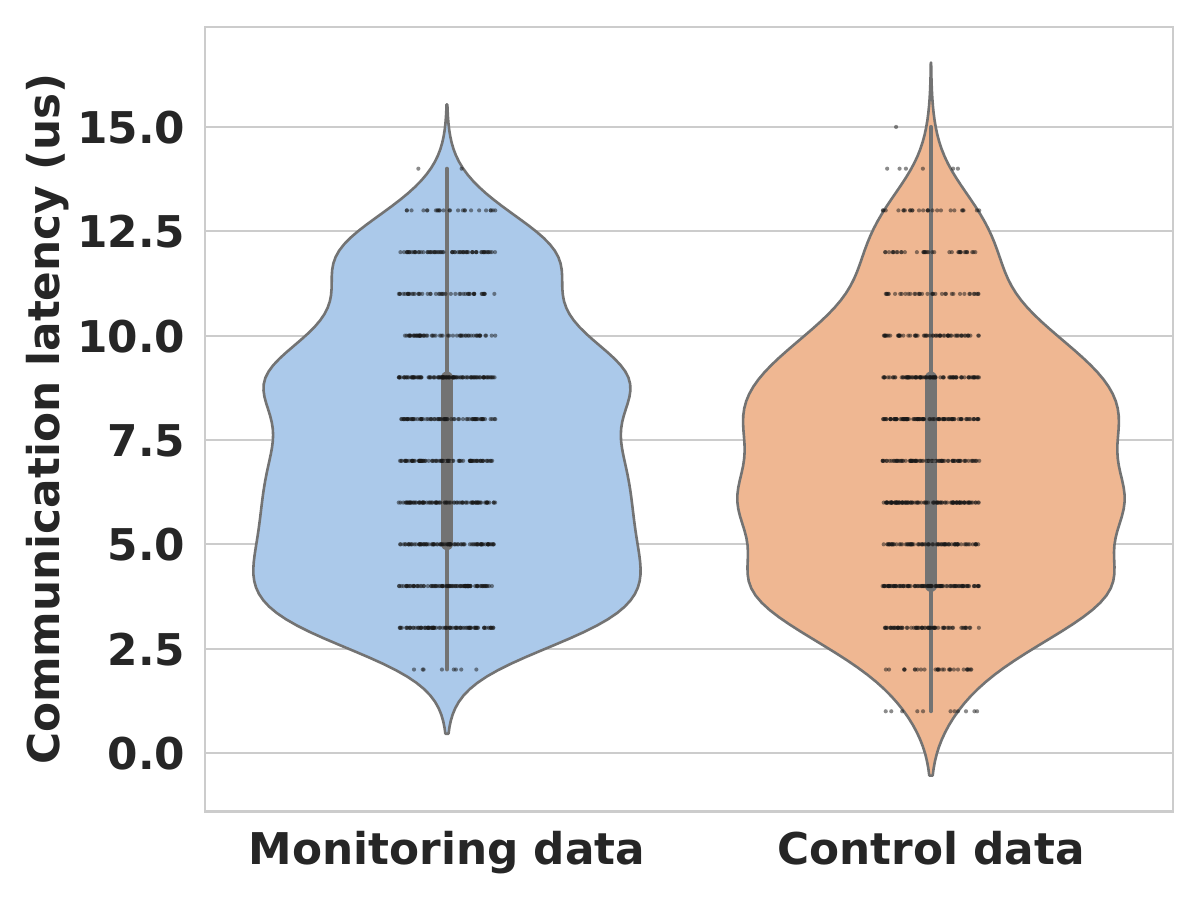}
	\caption{\bf Communication latency overhead for monitoring and control data to/from the AI processor runtime.}
	\label{fig:communication_overhead}
    \vspace{-2mm}
\end{figure}

To evaluate the efficiency of our far edge AI processor runtime under realistic conditions, we instrumented the popular open source RAN stack of srsRAN~\cite{srsran} with hooks that allow us to capture several important events across all the layers of the RAN (e.g., fronthaul packets, MAC scheduling events, PDCP packets, etc.), and to send back control decisions~\cite{jrtc-apps}. We deployed our custom stack and the AI processor runtime on an enterprise scale private 5G testbed~\cite{bahl2023accelerating}, and we measured the latency of delivering data between the programmable probes of srsRAN and the AI processor runtime. As illustrated in the violin plots of Figure~\ref{fig:communication_overhead}, in a well tuned system, the communication latency overhead in both the monitoring and control directions remains below 16$\mu$s, demonstrating the feasibility of our proposed AI processor runtime design. 

\subsection{Distributed AI orchestrator}
\label{sec:orch_eval}

To demonstrate the benefits of distributed AI orchestration, we developed a proof of concept orchestrator. The orchestrator is capable of dynamically orchestrating container-based applications across the edge and the cloud, taking into account their GPU and network requirements.

For our evaluation, we consider a setup composed of a far edge location with a single HPE DL110 Linux server equipped with an Nvidia L4 GPU, with 24GB of memory, and an Azure cloud VM equipped with an Nvidia A100 GPU with 40GB of memory. 
As sample workloads we use the SpotLight~\cite{sun2024spotlight} AI-based RAN anomlay detection and the visual SLAM of the Nvidia Isaac ROS framework~\cite{isaac_ros}.
The two applications were chosen as representative examples of AI-for-RAN and AI-on-RAN scenarios, and they both require a GPU.

By profiling the applications, we observe that SpotLight requires approximately 4\% of the edge GPU and 500Kbps of network bandwidth for performing anomaly detection for a single 5G cell, when using the dataset provided by~\cite{sun2024spotlight}.
In the case of the vSLAM, a single instance requires 8GB of GPU memory and $\sim$40Mbps of network bandwidth for an RGB-D camera configured with a resolution of 1280$\times$720 at 15 FPS.

\begin{figure}[t]
	\centering
	\includegraphics[width=0.9\columnwidth]{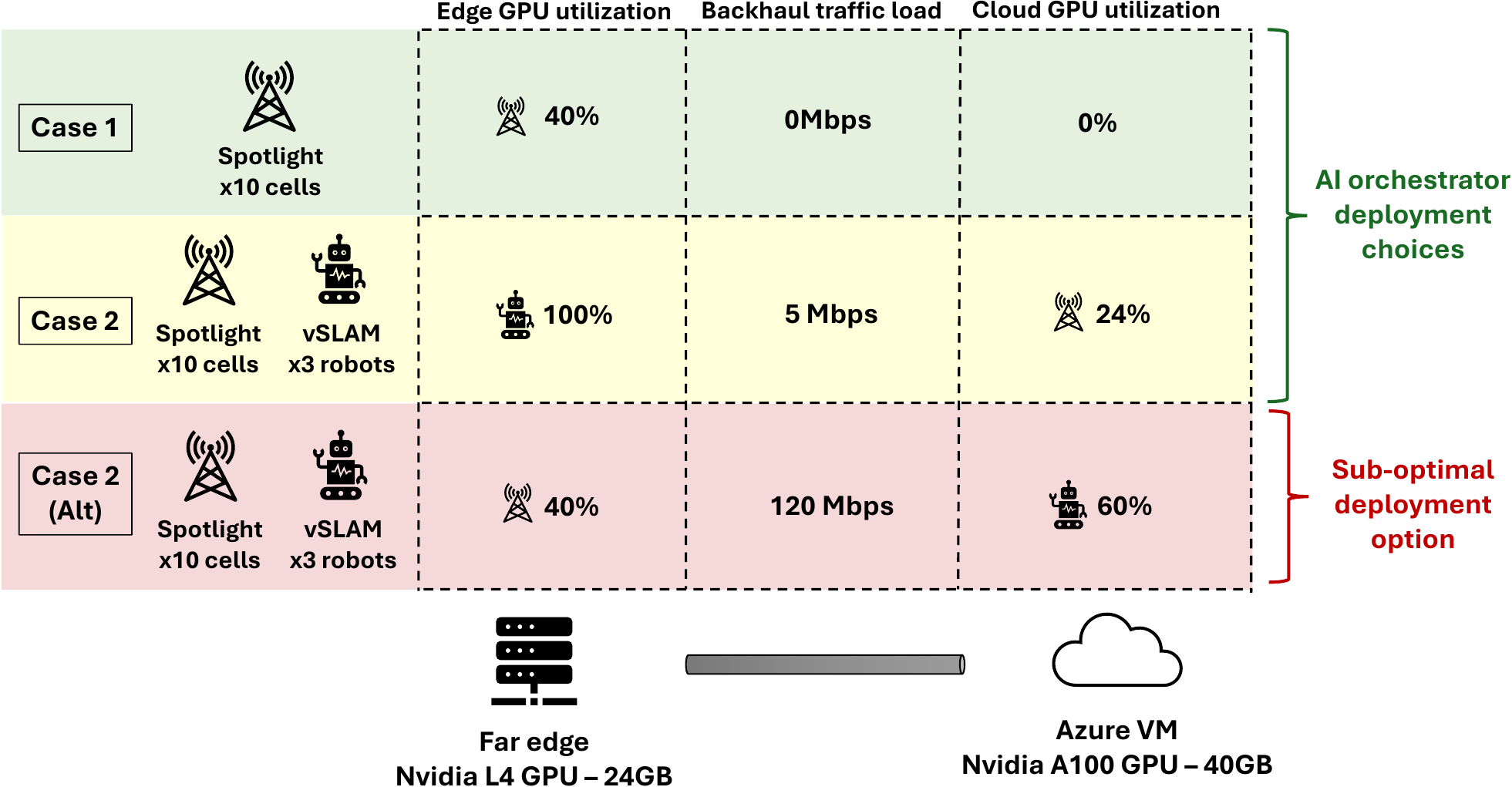}
	\caption{\bf AI orchestrator deployment choices vs sub-optimal option.}
	\label{fig:orchestration_eval}
    \vspace{-2mm}
\end{figure}

As illustrated in Figure~\ref{fig:orchestration_eval}, we deploy SpotLight across 10 cells (case 1). 
The AI orchestrator decides to orchestrate all the SpotLight instances at the edge location, considering that the resources of the L4 GPU are adequate (utilization of $\sim$40\%) and this deployment choice requires shipping no data to the cloud.
Then, we request to deploy three instances of the visual SLAM application.
The remaining edge GPU capacity is not adequate to accommodate the new workload, which requires 24GB of memory.
As such, the AI orchestrator migrates the SpotLight applications to the cloud and deploys the visual SLAM applications to the edge (case 2). 
The reason behind this deployment choice of the orchestrator is that it maximizes the utilization of the edge GPU (100\% utilization), while minimizing the amount of data that has to be shipped to the cloud. An alternative sub-optimal deployment option would be to deploy the visual SLAM to the cloud and keep SpotLight to the edge, however, this would lead to a requirement of shipping 120Mbps worth of data to the cloud (Case 2 - Alt). Along the lines of the discussion of Section~\ref{sec:platform}, the orchestrator could also be tailored to take into account other optimization parameters, such as the latency or the deployment cost.

\section{Open vs Closed architecture and interfaces for integration with RAN}

Despite the progress within the O-RAN community, many fundamental tensions remain around control interfaces. 
Several major vendors do not support the idea of a near-RT RIC~\cite{nrt_ric}, expressing concerns regarding the network stability and security. 
If RAN vendors allow developers to exert fine-grained control on the RAN behavior, these may clash with the proprietary algorithms vendors have implemented. 
Also, some vendors argue that exposing sensitive RAN data may affect the security of the network and the data privacy. 
Vendors are also reluctant to open up some of their interfaces, as this may reduce their competitive advantage. 
Our view is that the distributed AI platform proposed in this work should be a building block that can be customized appropriately for different use cases. 
We briefly discuss some examples of deployment options (Fig.~\ref{fig:model_split}).\looseness=-1

\begin{figure}[t]
	\centering
	\includegraphics[width=0.9\columnwidth]{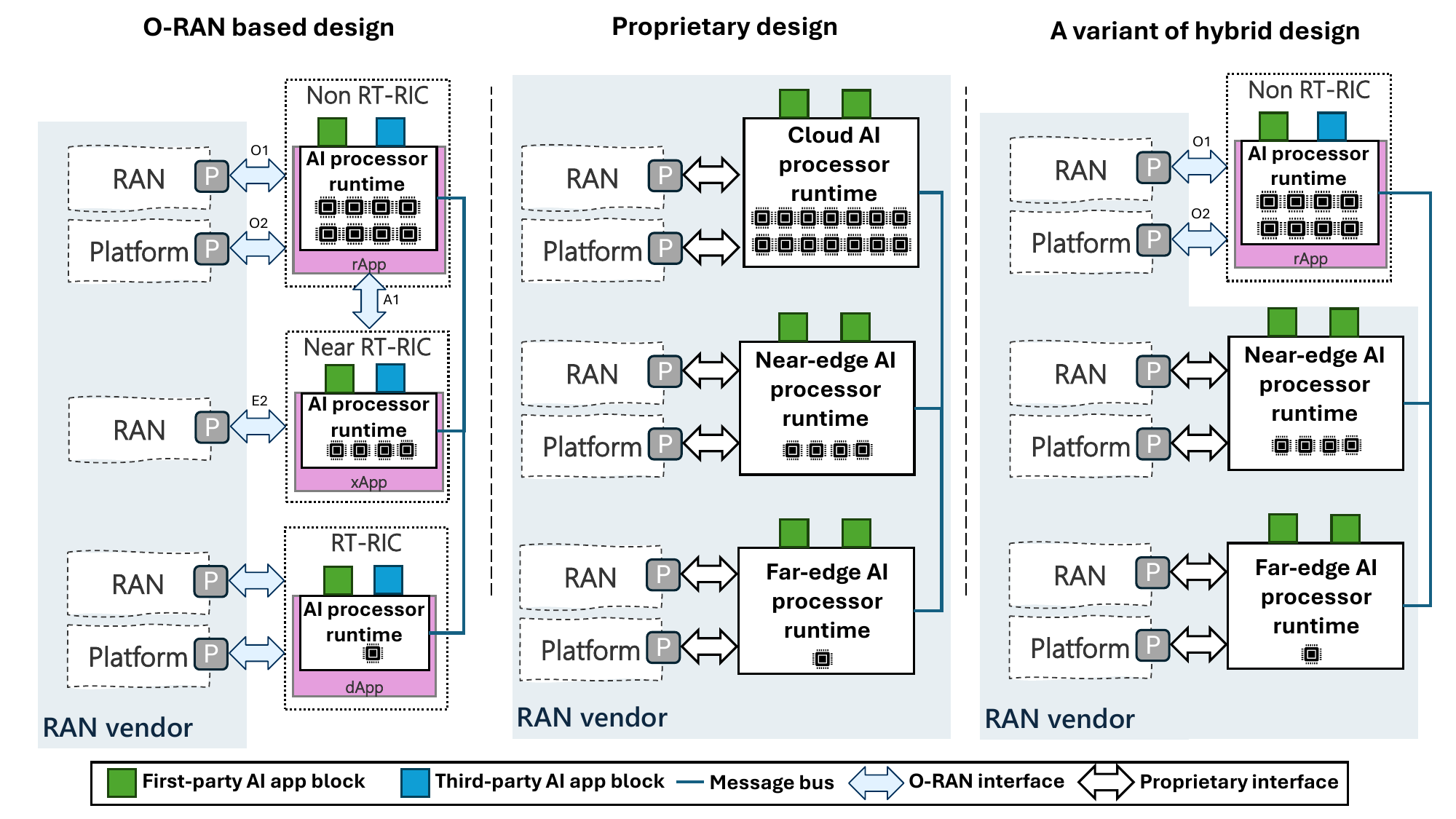}
	\caption{\bf Deployment options for integration of distributed AI platform with the RAN.}
	\label{fig:integration_options}
    \vspace{-2mm}
\end{figure}

\noindent\textbf{O-RAN based design --}
One extreme is an O-RAN based design (Fig.~\ref{fig:integration_options}, left), allowing any first or third-party vendor to deploy AI applications across the RAN infrastructure. 
One could leverage the O-RAN RICs and integrate the AI processor runtimes in them as apps (i.e., rApps, xApps, dApps).
For the local data collection and control operations, the AI processor runtimes could leverage the open RIC interfaces (e.g., E2, O1), augmented with programmable probes and exposed to the AI runtimes through appropriate adapters.
The communication between the AI processor runtimes could be facilitated by a message bus, which could be standardized and realized as an overlay network on top of the RIC fabric.
For the far-edge, and considering that there is currently no real-time RIC specification, one could leverage our far-edge AI processor runtime design as a blueprint.

\noindent\textbf{Proprietary design --}
Another extreme is a design, fully controlled by the RAN vendor (Fig.~\ref{fig:integration_options}, middle). 
In this case, the group that is in charge of the RAN product development could also be in charge of developing and managing the AI platform and of defining proprietary interfaces for data collection and control. 
A different group could deploy and manage their own AI RAN applications, without having to constantly go through the RAN product development group, to ask them to introduce new features or expose more data.
By decoupling the responsibilities, the vendor can accelerate its innovation, while still maintaining a full control over the RAN behavior. 

\noindent\textbf{Hybrid design --}
It is also possible to create a hybrid (Fig.~\ref{fig:integration_options}, right), that would be a mix of proprietary AI processor runtimes and runtimes running on top of standard O-RAN RICs.
AI applications could be composed of a mixture of first-party and third-party AI blocks, where the proprietary data and control interfaces of the RAN functions can only be accessed by the first-party AI blocks, whereas third-party blocks can only access APIs exposed by the standard interfaces.
This would allow the RAN vendor to maintain control over the RAN behavior, while still allowing third-party developers to innovate.\looseness=-1
 
\section{Conclusions}
To realize the 6G vision, we argued about the benefits of introducing AI in the RAN, from the management and infrastructure up to the application layer.
Based on these observations, we outlined the challenges for deploying AI solutions, stemming from the requirements of the applications and the characteristics of the RAN infrastructure.
Motivated by these challenges, we proposed a distributed AI platform architecture. 
Its goal is to alleviate the common painpoints of deploying AI applications in the RAN without being prescriptive about the implementation details. 
By identifying the future requirements and by initiating a discussion on the architecture of a 6G AI platform will help both the standards and the vendors to create opportunities for introducing AI solutions in 6G. 

\bibliographystyle{ACM-Reference-Format}
\bibliography{refs-1}


\begin{thebibliography}{27}


\ifx \showCODEN    \undefined \def \showCODEN     #1{\unskip}     \fi
\ifx \showDOI      \undefined \def \showDOI       #1{#1}\fi
\ifx \showISBNx    \undefined \def \showISBNx     #1{\unskip}     \fi
\ifx \showISBNxiii \undefined \def \showISBNxiii  #1{\unskip}     \fi
\ifx \showISSN     \undefined \def \showISSN      #1{\unskip}     \fi
\ifx \showLCCN     \undefined \def \showLCCN      #1{\unskip}     \fi
\ifx \shownote     \undefined \def \shownote      #1{#1}          \fi
\ifx \showarticletitle \undefined \def \showarticletitle #1{#1}   \fi
\ifx \showURL      \undefined \def \showURL       {\relax}        \fi
\providecommand\bibfield[2]{#2}
\providecommand\bibinfo[2]{#2}
\providecommand\natexlab[1]{#1}
\providecommand\showeprint[2][]{arXiv:#2}

\bibitem[Alliance(2024)]%
        {airan-alliance}
\bibfield{author}{\bibinfo{person}{AI-RAN Alliance}.} \bibinfo{year}{Aug 2024}\natexlab{}.
\newblock \bibinfo{booktitle}{\emph{{Integrating AI/ML in Open-RAN: Overcoming Challenges and Seizing Opportunities}}}.
\newblock \bibinfo{type}{{T}echnical {R}eport}. \bibinfo{institution}{AI-RAN Alliance}.
\newblock


\bibitem[Bahl et~al\mbox{.}(2023)]%
        {bahl2023accelerating}
\bibfield{author}{\bibinfo{person}{Paramvir Bahl}, \bibinfo{person}{Matthew Balkwill}, \bibinfo{person}{Xenofon Foukas}, \bibinfo{person}{Anuj Kalia}, \bibinfo{person}{Daehyeok Kim}, \bibinfo{person}{Manikanta Kotaru}, \bibinfo{person}{Zhihua Lai}, \bibinfo{person}{Sanjeev Mehrotra}, \bibinfo{person}{Bozidar Radunovic}, \bibinfo{person}{Stefan Saroiu}, {et~al\mbox{.}}} \bibinfo{year}{2023}\natexlab{}.
\newblock \showarticletitle{{Accelerating Open RAN Research Through an Enterprise-scale 5G Testbed}}. In \bibinfo{booktitle}{\emph{Proceedings of the 29th Annual International Conference on Mobile Computing and Networking}}. \bibinfo{pages}{1--3}.
\newblock


\bibitem[Balasingam et~al\mbox{.}(2024)]%
        {slicing}
\bibfield{author}{\bibinfo{person}{Arjun Balasingam} {et~al\mbox{.}}} \bibinfo{year}{2024}\natexlab{}.
\newblock \showarticletitle{{Application-Level} Service Assurance with 5G {RAN} Slicing}. In \bibinfo{booktitle}{\emph{USENIX NSDI}}.
\newblock


\bibitem[Bartsiokas et~al\mbox{.}(2022)]%
        {bartsiokas2022ml}
\bibfield{author}{\bibinfo{person}{Ioannis~A Bartsiokas} {et~al\mbox{.}}} \bibinfo{year}{2022}\natexlab{}.
\newblock \showarticletitle{{ML-based radio resource management in 5G and beyond networks: A survey}}.
\newblock \bibinfo{journal}{\emph{IEEE Access}}  \bibinfo{volume}{10} (\bibinfo{year}{2022}).
\newblock


\bibitem[Chorti et~al\mbox{.}(2022)]%
        {chorti2022context}
\bibfield{author}{\bibinfo{person}{Arsenia Chorti} {et~al\mbox{.}}} \bibinfo{year}{2022}\natexlab{}.
\newblock \showarticletitle{Context-aware security for 6G wireless: The role of physical layer security}.
\newblock \bibinfo{journal}{\emph{IEEE Communications Standards Magazine}} \bibinfo{volume}{6}, \bibinfo{number}{1} (\bibinfo{year}{2022}), \bibinfo{pages}{102--108}.
\newblock


\bibitem[D'Oro et~al\mbox{.}(2022)]%
        {d2022dapps}
\bibfield{author}{\bibinfo{person}{Salvatore D'Oro} {et~al\mbox{.}}} \bibinfo{year}{2022}\natexlab{}.
\newblock \showarticletitle{{dApps: Distributed applications for real-time inference and control in O-RAN}}.
\newblock \bibinfo{journal}{\emph{IEEE Communications}} (\bibinfo{year}{2022}).
\newblock


\bibitem[{eBPF}({[n.\,d.]})]%
        {ebpf}
\bibfield{author}{\bibinfo{person}{{eBPF}}.} \bibinfo{year}{[n.\,d.]}\natexlab{}.
\newblock \bibinfo{title}{{Dynamically program the kernel for efficient networking, observability, tracing, and security}}.
\newblock
\newblock
\newblock
\shownote{\url{https://ebpf.io/}}.


\bibitem[Foukas et~al\mbox{.}(2023)]%
        {foukas2023taking}
\bibfield{author}{\bibinfo{person}{Xenofon Foukas} {et~al\mbox{.}}} \bibinfo{year}{2023}\natexlab{}.
\newblock \showarticletitle{Taking 5G RAN analytics and control to a new level}. In \bibinfo{booktitle}{\emph{ACM MobiCom}}. \bibinfo{pages}{1--16}.
\newblock


\bibitem[Foukas et~al\mbox{.}(2017)]%
        {foukas2017orion}
\bibfield{author}{\bibinfo{person}{Xenofon Foukas}, \bibinfo{person}{Mahesh~K Marina}, {and} \bibinfo{person}{Kimon Kontovasilis}.} \bibinfo{year}{2017}\natexlab{}.
\newblock \showarticletitle{Orion: RAN slicing for a flexible and cost-effective multi-service mobile network architecture}. In \bibinfo{booktitle}{\emph{Proceedings of the 23rd annual international conference on mobile computing and networking}}. \bibinfo{pages}{127--140}.
\newblock


\bibitem[Foukas and Radunovic(2021)]%
        {foukas2021concordia}
\bibfield{author}{\bibinfo{person}{Xenofon Foukas} {and} \bibinfo{person}{Bozidar Radunovic}.} \bibinfo{year}{2021}\natexlab{}.
\newblock \showarticletitle{Concordia: Teaching the 5G vRAN to share compute}. In \bibinfo{booktitle}{\emph{ACM SIGCOMM}}.
\newblock


\bibitem[Foukas and Radunovic(2025)]%
        {foukas2025future}
\bibfield{author}{\bibinfo{person}{Xenofon Foukas} {and} \bibinfo{person}{Bozidar Radunovic}.} \bibinfo{year}{2025}\natexlab{}.
\newblock \showarticletitle{The future of the industrial AI edge is cellular}. In \bibinfo{booktitle}{\emph{Proceedings of the 26th International Workshop on Mobile Computing Systems and Applications}}. \bibinfo{pages}{61--66}.
\newblock


\bibitem[Kalia et~al\mbox{.}(2025)]%
        {kalia2025towards}
\bibfield{author}{\bibinfo{person}{Anuj Kalia} {et~al\mbox{.}}} \bibinfo{year}{2025}\natexlab{}.
\newblock \showarticletitle{Towards Energy Efficient 5G vRAN Servers}. In \bibinfo{booktitle}{\emph{USENIX Symposium on Networked Systems Design and Implementation (NSDI)}}.
\newblock


\bibitem[Ko et~al\mbox{.}(2024)]%
        {ko2024edgeric}
\bibfield{author}{\bibinfo{person}{Woo-Hyun Ko}, \bibinfo{person}{Ushasi Ghosh}, \bibinfo{person}{Ujwal Dinesha}, \bibinfo{person}{Raini Wu}, \bibinfo{person}{Srinivas Shakkottai}, {and} \bibinfo{person}{Dinesh Bharadia}.} \bibinfo{year}{2024}\natexlab{}.
\newblock \showarticletitle{$\{$EdgeRIC$\}$: Empowering real-time intelligent optimization and control in $\{$NextG$\}$ cellular networks}. In \bibinfo{booktitle}{\emph{21st USENIX Symposium on Networked Systems Design and Implementation (NSDI 24)}}. \bibinfo{pages}{1315--1330}.
\newblock


\bibitem[Kundu et~al\mbox{.}(2024)]%
        {kundu2024towards}
\bibfield{author}{\bibinfo{person}{Lopamudra Kundu} {et~al\mbox{.}}} \bibinfo{year}{2024}\natexlab{}.
\newblock \showarticletitle{{Towards Energy Efficient RAN: From Industry Standards to Trending Practice}}.
\newblock \bibinfo{journal}{\emph{arXiv:2402.11993}} (\bibinfo{year}{2024}).
\newblock


\bibitem[{Microsoft}({[n.\,d.]})]%
        {jbpf}
\bibfield{author}{\bibinfo{person}{{Microsoft}}.} \bibinfo{year}{[n.\,d.]}\natexlab{}.
\newblock \bibinfo{title}{{Userspace instrumentation and control framework for deploying control and monitoring functions in a secure manner}}.
\newblock
\newblock
\newblock
\shownote{\url{https://github.com/microsoft/jbpf}}.


\bibitem[{Microsoft}(2025a)]%
        {jrtcontroller}
\bibfield{author}{\bibinfo{person}{{Microsoft}}.} \bibinfo{year}{2025}\natexlab{a}.
\newblock \bibinfo{title}{jrt-controller}.
\newblock \bibinfo{howpublished}{\url{https://github.com/microsoft/jrt-controller}}.
\newblock


\bibitem[{Microsoft}(2025b)]%
        {jrtc-apps}
\bibfield{author}{\bibinfo{person}{{Microsoft}}.} \bibinfo{year}{2025}\natexlab{b}.
\newblock \bibinfo{title}{jrtc-apps}.
\newblock \bibinfo{howpublished}{\url{https://github.com/microsoft/jrtc-apps}}.
\newblock


\bibitem[Morris(2024)]%
        {nrt_ric}
\bibfield{author}{\bibinfo{person}{Iain Morris}.} \bibinfo{year}{2024}\natexlab{}.
\newblock \bibinfo{title}{{AT\&T, all in with Ericsson, seems to have shut the door to xApps}}.
\newblock
\newblock
\newblock
\shownote{\url{https://bit.ly/3TZzYP0}}.


\bibitem[{Nvidia}({[n.\,d.]})]%
        {isaac_ros}
\bibfield{author}{\bibinfo{person}{{Nvidia}}.} \bibinfo{year}{[n.\,d.]}\natexlab{}.
\newblock \bibinfo{title}{{NVIDIA Isaac ROS}}.
\newblock
\newblock
\newblock
\shownote{\url{https://developer.nvidia.com/isaac/ros}}.


\bibitem[Nvidia(2023)]%
        {nvidia_vran}
\bibfield{author}{\bibinfo{person}{Nvidia}.} \bibinfo{year}{2023}\natexlab{}.
\newblock \bibinfo{title}{{Building Software-Defined, High-Performance, and Efficient vRAN}}.
\newblock
\newblock
\newblock
\shownote{\url{http://bit.ly/4mjBiIF}}.


\bibitem[Schiavo et~al\mbox{.}(2024)]%
        {schiavo2024yinyangran}
\bibfield{author}{\bibinfo{person}{Leonardo~Lo Schiavo} {et~al\mbox{.}}} \bibinfo{year}{2024}\natexlab{}.
\newblock \showarticletitle{YinYangRAN: Resource Multiplexing in GPU-Accelerated Virtualized RANs}. In \bibinfo{booktitle}{\emph{IEEE INFOCOM}}. \bibinfo{pages}{1--10}.
\newblock


\bibitem[Singh et~al\mbox{.}(2021)]%
        {singh2021energy}
\bibfield{author}{\bibinfo{person}{Rajkarn Singh} {et~al\mbox{.}}} \bibinfo{year}{2021}\natexlab{}.
\newblock \showarticletitle{Energy-efficient orchestration of metro-scale 5G radio access networks}. In \bibinfo{booktitle}{\emph{IEEE INFOCOM 2021-IEEE Conference on Computer Communications}}. IEEE, \bibinfo{pages}{1--10}.
\newblock


\bibitem[Soldani et~al\mbox{.}(2023)]%
        {soldani2023ebpf}
\bibfield{author}{\bibinfo{person}{David Soldani} {et~al\mbox{.}}} \bibinfo{year}{2023}\natexlab{}.
\newblock \showarticletitle{ebpf: A new approach to cloud-native observability, networking and security for current (5g) and future mobile networks (6g and beyond)}.
\newblock \bibinfo{journal}{\emph{IEEE Access}}  \bibinfo{volume}{11} (\bibinfo{year}{2023}), \bibinfo{pages}{57174--57202}.
\newblock


\bibitem[{srsRAN}({[n.\,d.]})]%
        {srsran}
\bibfield{author}{\bibinfo{person}{{srsRAN}}.} \bibinfo{year}{[n.\,d.]}\natexlab{}.
\newblock \bibinfo{title}{{srsRAN Project -- Open Source RAN}}.
\newblock
\newblock
\newblock
\shownote{\url{https://www.srslte.com/}}.


\bibitem[Sun et~al\mbox{.}(2024)]%
        {sun2024spotlight}
\bibfield{author}{\bibinfo{person}{Chuanhao Sun} {et~al\mbox{.}}} \bibinfo{year}{2024}\natexlab{}.
\newblock \showarticletitle{{SpotLight: Accurate, Explainable and Efficient Anomaly Detection for Open RAN}}. In \bibinfo{booktitle}{\emph{ACM MobiCom}}.
\newblock


\bibitem[Trevlakis et~al\mbox{.}(2023)]%
        {trevlakis2023localization}
\bibfield{author}{\bibinfo{person}{Stylianos~E Trevlakis} {et~al\mbox{.}}} \bibinfo{year}{2023}\natexlab{}.
\newblock \showarticletitle{Localization as a key enabler of 6G wireless systems: A comprehensive survey and an outlook}.
\newblock \bibinfo{journal}{\emph{IEEE Open Journal of the Communications Society}} (\bibinfo{year}{2023}).
\newblock


\bibitem[Zhang et~al\mbox{.}(2024)]%
        {zhang2024vulcan}
\bibfield{author}{\bibinfo{person}{Yiwen Zhang} {et~al\mbox{.}}} \bibinfo{year}{2024}\natexlab{}.
\newblock \showarticletitle{{Vulcan: Automatic Query Planning for Live ML Analytics}}. In \bibinfo{booktitle}{\emph{USENIX NSDI}}.
\newblock


\end{thebibliography}

\end{document}